\documentclass[journal]{IEEEtran}
\usepackage{caption}
\usepackage{subcaption}
\usepackage{mathrsfs}
\usepackage[cmex10]{amsmath}
\usepackage{cite,graphicx,epsfig,wrapfig,amsfonts,amssymb,algorithmic,algorithm,threeparttable,color,url}
\usepackage{mdwmath}
\usepackage{mdwtab}
\usepackage{amsthm}

\theoremstyle{definition}

\graphicspath{{./}{figures/}}

\ifCLASSINFOpdf
\else
\fi
\begin{document}

\begin{titlepage}
\begin{center}
\vspace*{-2\baselineskip}
\begin{minipage}[l]{7cm}
\flushleft
\includegraphics[width=2 in]{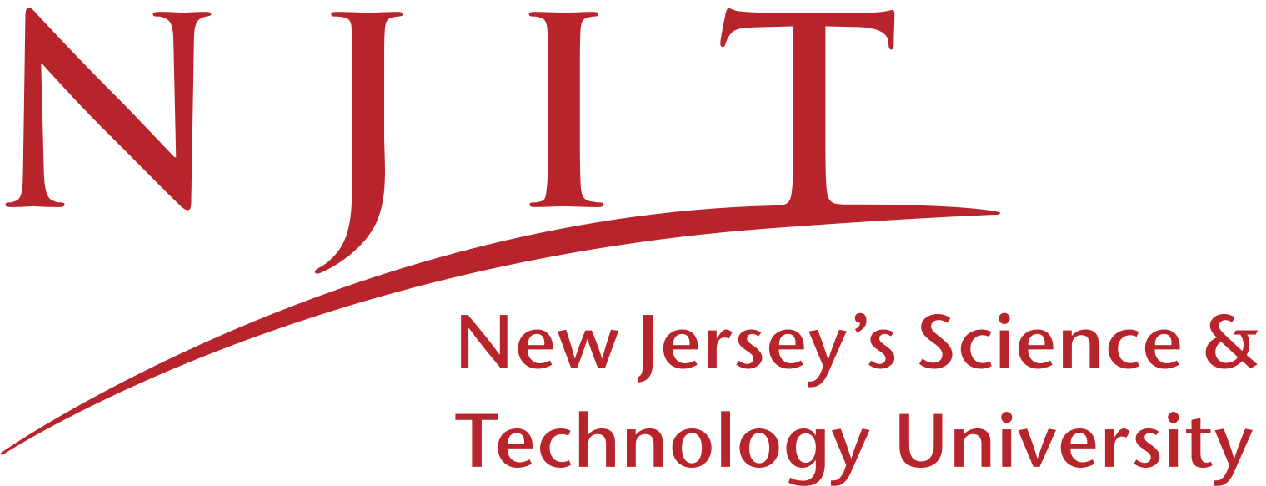}
\end{minipage}
\hfill
\begin{minipage}[r]{7cm}
\flushright
\includegraphics[width=1 in]{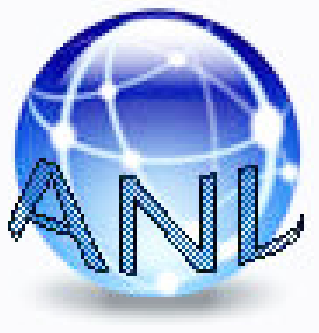}%
\end{minipage}

\vfill

\textsc{\LARGE FreeNet: Spectrum and Energy Harvesting Wireless Networks}\\

\vfill
\textsc{\LARGE Nirwan Ansari\\[12pt]
\LARGE Tao Han}\\
\vfill
\textsc{\LARGE TR-ANL-2015-001\\[12pt]
\LARGE March 16, 2015}\\[1.5cm]
\vfill
{ADVANCED NETWORKING LABORATORY\\
 DEPARTMENT OF ELECTRICAL AND COMPUTER ENGINEERING\\
 NEW JERSY INSTITUTE OF TECHNOLOGY}
\end{center}
\end{titlepage}

\title{FreeNet: Spectrum and Energy Harvesting Wireless Networks}
\author{\IEEEauthorblockN{Nirwan Ansari, \emph{Fellow, IEEE}, and Tao Han, \emph{Student Member, IEEE}
}\\
\IEEEauthorblockA{Advanced Networking Laboratory \\
Department of Electrical and Computer Engineering \\
New Jersey Institute of Technology, Newark, NJ, 07102, USA\\
Email:  \{nirwan.ansari,th36\}@njit.edu}

\thanks{This work was supported in part by NSF under grant no. CNS-1320468, FreeNet: Cognitive Wireless Networking Powered by Green Energy.}}
\maketitle

\begin{abstract}
The dramatic mobile data traffic growth is not only resulting in the spectrum crunch but is also leading to exorbitant energy consumption. It is thus desirable to liberate mobile and wireless networks from the constraint of the spectrum scarcity and to rein in the growing energy consumption. This article introduces FreeNet, figuratively synonymous to ``Free Network", which engineers the spectrum and energy harvesting techniques to alleviate the spectrum and energy constraints by sensing and harvesting spare spectrum for data communications and utilizing renewable energy as power supplies, respectively. Hence, FreeNet increases the spectrum and energy efficiency of wireless networks and enhances the network availability. As a result, FreeNet can be deployed to alleviate network congestion in urban areas, provision broadband services in rural areas, and upgrade emergency communication capacity. This article provides a brief analysis of the design of FreeNet that accommodates the dynamics of the spare spectrum and employs renewable energy.
\end{abstract}
\IEEEpeerreviewmaketitle

\section{Introduction}
Proliferation of wireless devices drive the exponential growth of wireless data traffic over wireless and mobile networks. This leads to a continuous surge not only in network capacity demands but also in network energy consumption. Since a wireless system is spectrum limited, the ever-increasing capacity demands result in the spectrum crunch. Deploying additional network infrastructures, e.g., base stations (BSs), is an effective approach to alleviate the spectrum shortage \cite{Andrews:2014:AOLB}. Thus, small cell BSs (SCBSs) will be widely deployed. SCBSs can provide high network capacity for wireless users by capitalizing on their close proximity to the users. However, a SCBS has a limited coverage area. Thus, the number of SCBSs will be orders of magnitude larger than that of macro BSs (MBSs) for a large scale network deployment. As a result, the overall energy consumption of cellular networks will keep increasing \cite{Han:2012:OGC}. Therefore, current wireless access networks are eventually strained by the spectrum scarcity and energy consumption. It is desirable to amalgamate the spectrum harvesting and energy harvesting technologies to liberate the wireless networks from these constraints.

\subsection{Spectrum Harvesting Techniques}
Spectrum harvesting techniques explore the under-utilized licensed spectrum to boost the capacity of wireless and mobile networks and enable dynamic access, allocation and aggregation of licensed, unlicensed and TV white space frequency channels. Spectrum harvesting techniques consist of three major functions:
\begin{itemize}
\item\textit{Spectrum Sensing \cite{Yucek:2009:SSS}:} one of the essential functions of spectrum sensing is to identify the spectrum opportunities, e.g., spectrum holes. The identified spectrum holes can be utilized by the secondary transceivers without interfering the primary users\footnote{The primary users refer to the users who own the license to the spectrum.}. The other function of spectrum sensing is monitoring the primary users' activities on using the spectrum holes. When primary users resume their transmissions over the spectrum holes, the secondary users should evacuate from the spectrum holes to avoid the interference.
\item\textit{Spectrum Management \cite{Akyildiz:2008:SSM}:} this function manages the spectrum sensing and access strategies. The spectrum sensing strategies specify when and which spectrum to sense. Based on the spectrum sensing results, the spectrum access strategies evaluate the spectrum holes and decide whether they are accessible.
\item\textit{Spectrum Sharing \cite{Akyildiz:2008:SSM}:} this function coordinates the spectrum allocation among users to mitigate the inference and collisions among the users. Moreover, this function aggregates available frequency channels according to the users' capacity demands.
\end{itemize}

\subsection{Energy Harvesting Techniques}
Energy harvesting techniques enable network devices, e.g., BSs, to generate electricity from renewable energy sources such as wind, solar, and sustainable biofuels. Continuous advances in energy harvesting technologies are improving the efficiency of electricity generation and reducing the cost of energy harvesting systems. Therefore, energy harvesting techniques, by exploiting renewable energy, are promising to alleviate the power consumption of wireless networks \cite{Han:2014:PMN}.

\begin{figure}[t]
\centering
\includegraphics[scale=0.25]{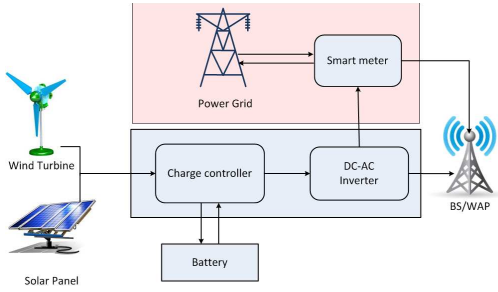}
\caption{A energy harvesting BS.}
\label{fig:green_bs}
\end{figure}

As shown in Fig. \ref{fig:green_bs}, energy harvesting BSs may consist of several power related components, which are the renewable power generators, e.g., solar panels and wind turbines, the charge controller that regulates the output voltage of each power generator, the DC-AC inverter that changes direct current (DC) to alternating current (AC), and the battery that stores surplus power. If the energy harvesting BSs are also connected to the power grid, the BS may equip with a smart meter that enables the power transmission between BSs and the power grid. In addition, multiple BSs may share an energy harvester, e.g., a wind farm, for power supplies.

\subsection{FreeNet: Spectrum and Energy Harvesting Wireless Networks}
\begin{figure}[h]
\centering
\includegraphics[scale=0.35]{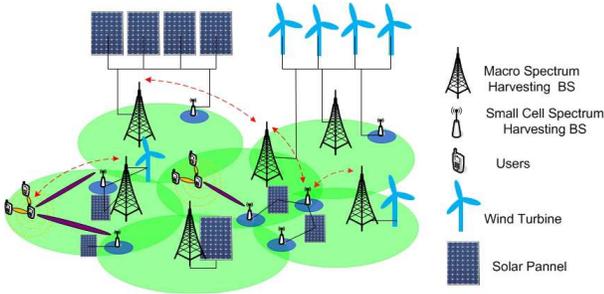}
\caption{FreeNet: spectrum and energy harvesting wireless network.}
\label{fig:futureaccess}
\end{figure}

By leveraging spectrum and energy harvesting technologies, FreeNet typically comprises various spectrum harvesting BSs powered by renewable energy from the energy harvesters, as shown in Fig. \ref{fig:futureaccess}. These spectrum harvesting BSs are mainly powered by renewable energy while the power grid is used as the back-up.
On the one hand, by taking advantage of spectrum harvesting technologies, FreeNet liberates the wireless network from the spectrum constraint, in which wireless nodes sense and utilize the available spectrum for data communications. In this way, FreeNet enhances the spectrum agility and energy efficiency of the network because wireless nodes are able to sense and utilize the spectrum over which they experience less fading and interference, improves network ubiquity because wireless nodes are able to access the network as long as they can detect the available spectrum, and easily incorporates new technologies as long as rules on sensing and utilizing the spectrum are followed. On the other hand, envisioning renewable energy as an important energy resource for future wireless access networks, FreeNet explores the usage of renewable energy that is sustainable. Hence, FreeNet reduces the grid power consumption, and thus enhances the network energy efficiency as well as reduces carbon footprints. Moreover, FreeNet, by making use of renewable energy, can be featured as a drop-and-play network that improves the network ubiquity: FreeNet is able to provide data communications in areas without power grid such as remote rural areas or areas affected by natural disasters.

To provision high capacity wireless networks, FreeNet exploits the heterogeneity of wireless access networks and integrates both the infrastructure based and mesh-based access networks to provide network access to users. Moreover, FreeNet adopts advanced communication technologies such as bandwidth aggregation and smart antenna to enhance the spectrum and energy efficiency of the network. Based on the availability of the spectrum, renewable energy, and traffic demands, FreeNet adapts its operation strategies in terms of network topology, operation bandwidth, transmission techniques, and communication protocols. For infrastructure-based access networks, the network capacity is adapted by changing transmission techniques and coverage area of BSs. For the mesh-based access network, the network topology is adapted by optimizing routing algorithms.

FreeNet aims to build a self-organized and self-optimized wireless network in which the renewable energy powered network nodes can dynamically utilize the available spectrum for their data communications. The major concern of energy harvesting cognitive radio networks is to design spectrum sensing and access strategies for secondary network nodes that are powered by harvested energy \cite{Lu:2014:DSARF}. In comparison, FreeNet, though has to address the same issue, is more advanced and enables a higher order of freedom in terms of spectrum and energy usage and realizes a self-organized and self-optimized wireless network.

\section{FreeNet Application Scenarios}
In FreeNet, wireless nodes are able to dynamically access the available spectrum to enhance the spectrum efficiency of the network. In addition, wireless nodes are able to utilize the renewable energy as power supplies. In this way, FreeNet liberates the network from the constraints of the spectrum and energy. FreeNet will be deployed for three major purposes: 1) alleviating network congestion in urban areas, 2) provisioning broadband services in rural areas, and 3) enhancing the emergency communications capability in unexpected critical situations caused by natural disasters, such as hurricanes and earthquakes, or by terrorist attacks, such as the 9/11 attack.
\subsection{Alleviating Network Congestion in Urban Areas}
With the rapid development of radio access techniques and mobile devices, an increasing number of Internet applications are carried over mobile and wireless networks. Urban areas are characterized by high density of population, and thus demand high capacity wireless networks. Tremendous research efforts have been made to enhance the capacity and quality of service (QoS) of wireless access networks \cite{Han:2013:OAC}. FreeNet, which dynamically exploits the under utilized spectrum to enhance spectrum efficiency of wireless networks, can potentially alleviate network congestion in urban areas. Moreover, without the strict constraint of power supplies, the deploying locations of wireless access points can be optimized based on traffic demands and spectrum availability instead of being restricted by the power outlets \cite{Siomina:2006:AOS}.

In urban areas, FreeNet adopts heterogeneous networking architecture consisting of both macro BSs and small cell BSs. The BSs are able to harvest spectrum and utilize renewable energy. Based on the traffic demands and their geological distribution, traffic loads are balanced among the BSs according to the renewable energy status and the availability of spectrum. In addition, mesh based networks, e.g., device-to-device (D2D) communication networks, may be dynamically formed to further enhance the capacity and efficiency of the networks \cite{Doppler:2009:D2D}.

\subsection{Provisioning Broadband Services in Rural Areas}
Fibre or xDSL access may not be available in rural areas. To provide connectivity to communities in mountainous, undulating and remote terrains, WindFi base stations are designed to provide basic network connectivity in rural areas \cite{WindFi:2012}. WindFi base stations operate entirely on renewable energy and wireless access options for users are realized via point to point (P2P) hopping backhaul links and point to multi-point (P2MP) customer last mile access. However, the traffic demand from rural areas is also increasing. WindFi, with static frequency planning, may not be able to efficiently provision high network capacity. Since FreeNet is a drop-and-play network, multiple infrastructure-based energy harvesting BSs can be deployed to provide broadband services in remote rural areas. These BSs may dynamically utilize the available spectrum to provide users the network access as well as to work as the backhaul network. In addition, multiple mesh access networks can be formed in FreeNet to provide users with alternative network accesses. FreeNet, by optimizing the utilization of the available spectrum and renewable energy, will significantly increase the quality of the broadband services in remote rural areas.
\subsection{Enhancing Emergency Communication Capability in Critical situations}
Emergency communications refers to providing users with network accesses in unexpected critical situations \cite{Ansari:2008:NCC}. Usually, the destruction or extremely limited availability of the communications infrastructure of the region in distress exacerbates the rescue and recovery operations. Voice services, used to communicate among responders and with headquarters for control and command, or used by those affected by the emergency situation, may be severely restricted and unreliable even when available. Furthermore, it is often impossible to share and use all the relief resources through advanced information technologies--such as accessing remote databases, Web sites, and Web-based applications--and exchange data with agency headquarters and other field command centers.
Without the constraints of the radio frequency and energy supplies, FreeNet is a viable solution for provisioning emergency communication capability in these critical situations. Since the major goal of emergency communications is to disseminate surveillance information, FreeNet is able to provide users with basic communication connectivity, and extend lifetime and coverage area of the access network.

\section{Dynamic Network Architecture Optimization}
FreeNet consists of both infrastructure based wireless access networks and mesh-based wireless access networks.  To optimize the utilization of FreeNet, the network architecture should be adapted according to the spectrum opportunities, the availability of renewable energy, and the traffic load distribution.

In order to provision high network capacity, mobile and wireless access networks have adopted several advanced techniques such as the bandwidth aggregation, the smart antenna system, and small cell networks. The bandwidth aggregation technique is able to extend the bandwidth and thus enhance the network capacity. However, as the bandwidth increases, the transmit power should also be increased to keep the  transmit power per unit bandwidth constant. The smart antenna system enhances the capacity of wireless access networks by increasing the signal-to-interference-and-noise ratio (SINR). The multiple input multiple output (MIMO) system is one of the smart antenna systems that will be widely adopted in future wireless access networks. MIMO applies multiple antennas at both the transmitter and receiver to mitigate the channel fading, and thus increases the access network capacity. Smart antenna systems consume extra power since multiple transmit antennas are employed. Each additional antenna consumes extra DC power, which is constant irrespective of the transmission rate in the circuit, such as Analog to Digital Converter (ADC) and Digital to Analog Converter (DAC). The additional DC power consumption will affect the energy consumption of smart antenna systems. Small cell networks deploy low power BSs with close proximity to users. Because of the proximity, small cell BSs experience much less fading than macro BSs do. Therefore, given the same bandwidth and transmission power, a small cell BS can provide a higher network capacity. However, owing to the low transmission power, the coverage area of a small cell BS is limited. As a result, a large number of small cell BSs have to be deployed in order to cover the service area. The densely deployed small cell BSs may interfere with each other, and subsequently reduce the achievable network capacity of individual BSs.

FreeNet adapts its capacity by using the above methods according to the dynamics of the availability of the spectrum and renewable energy, and traffic demands.
For infrastructure-based access networks, BSs with higher renewable energy may adopt advanced transmission techniques and aggregate more bandwidth to serve more traffic demands. Since FreeNet also integrates mesh-based wireless access networks, the routing strategies are optimized to maximize the utilization of available spectrum and renewable energy while reducing grid power consumption. Badawy \emph{et al.} \cite{Badawy:2010:EPS} studied the energy aware routing in renewable energy powered mesh networks. By exploring the spectrum harvesting techniques, the routing strategies in FreeNet should also be aware of the available spectrum.

The availability of spectrum and renewable energy is highly dynamic. Since the optimal network architecture for FreeNet depends on the availability of spectrum and renewable energy, the optimal network architecture changes dynamically. However, adapting network architecture requires wireless nodes to negotiate and exchange operation parameters such as cell sizes and routing information; this introduces additional overhead to the access networks. Thus, adapting the network architecture frequently may result in excessive overheads and reduce the network efficiency. Therefore, the tradeoff between obtaining the optimal network architecture and the overhead incurred during the network architecture adaptions should be determined. To obtain optimal control of the network architecture adaptation, the inherent relationships among the availability of spectrum, the generation of renewable energy, and the data traffic demand should be investigated by using advanced probability theory, queuing theory, and machine learning techniques.

\section{Optimal Network Resource Management}
Since FreeNet adapts its network architecture according to the dynamics of the available spectrum and renewable energy, network resource allocation and management should be optimized under each network architecture to enhance the spectrum and energy efficiency of the network. FreeNet adopts spectrum and energy harvesting techniques. Thus, both spectrum and renewable energy are considered as network resources. It is desired to optimize the joint allocation of the spectrum and harvested energy to enhance the spectrum and energy efficiency of the network.
However, it is challenging to harness the spectrum dynamics and the renewable energy dynamics simultaneously. Thus, we translate the joint spectrum and renewable energy allocation problem into two coupled subproblems. The first one is to optimize the spectrum usage with the given renewable energy dynamics while the other is to optimize the renewable energy usage with the given spectrum dynamics.

\subsection{Spectrum Sensing and Sharing:}
FreeNet relies on spectrum harvesting techniques to dynamically utilize the spectrum. Thus, spectrum sensing and sharing is essential in FreeNet. Different from traditional cognitive radio networks, FreeNet is powered by renewable energy. Therefore, the spectrum sensing and sharing scheme for FreeNet should be optimized according to the dynamics of renewable energy. The spectrum sensing problem consists of two sub-problems: the first one is to form coalitions among wireless nodes to enhance the probability of sensing the available spectrum; the other problem is to decide which part of the spectrum each coalition should sense. The spectrum sensing problem can be formulated as:

\noindent\emph{Given:}
\begin{itemize}
\item The topology of a wireless access network.
\item The dynamics of renewable energy.
\end{itemize}
\noindent\emph{Obtain:} The optimal spectrum sensing strategies.

\noindent\emph{Objective:} To maximize the total amount of the sensed spectrum with minimal grid power consumption.

There are three major spectrum sharing strategies: spatial spectrum sharing, temporal spectrum sharing, and hybrid spectrum sharing \cite{Huang:2014:OGEP}. These solutions do not consider the renewable energy status in optimizing the spectrum sharing strategies, which may lead to an inefficient spectrum allocation. For example, a wireless access point with very limited amount of green energy may be allocated a large amount of spectrum that cannot be fully utilized because of the energy shortage. Thus, in FreeNet, the spectrum sharing strategies should be aware of the renewable energy status. Assuming that the nodes always have data to transmit, the spectrum sharing problem can be formulated as:

\noindent\emph{Given:}
\begin{itemize}
\item The available spectrum.
\item Interference relationships among different wireless nodes.
\item The dynamics of renewable energy.
\end{itemize}
\noindent\emph{Obtain:} The spectrum allocation among the wireless nodes.

\noindent\emph{Objective:} To maximize the network capacity.

\noindent\emph{Subject to:} The data rate requirements of individual nodes are satisfied.

\subsection{Renewable Energy Sharing}
Renewable energy as an important network resource for FreeNet should be optimized to minimize the grid power consumption of the network. Two scenarios of renewable energy sharing in FreeNet should be investigated: \begin{enumerate}
\item Multiple wireless nodes share the same renewable energy generator: in this scenario, renewable energy is shared among wireless nodes attached to the renewable energy generator. The objective of renewable energy sharing is to maximize the network capacity with the minimal grid power consumption \cite{Han:2014:PMN}.
\item Individual wireless nodes exclusively use their own renewable energy generators: in this case, renewable energy cannot be shared among wireless access points directly. However, the power consumption among wireless access points can be balanced by adapting the coverage area of individual wireless access points \cite{Han:2013:OOG}, in which the wireless access points with higher amount of renewable energy increase their coverage areas to absorb more traffic. In this way, the probability that the network consumes grid power is reduced, and the utilization of renewable energy is improved.
In addition to cell size adaptation, wireless access points with higher amount of renewable energy may activate advanced transmission techniques such as MIMO and transmission beamforming to enhance the spectrum efficiency. Therefore, these wireless access points can satisfy their users' traffic demands with less spectrum. The saved spectrum can be allocated to the wireless access points which are short of renewable energy. Given the users' traffic demands, increasing the bandwidth allocation toward a wireless access point reduces the wireless access point's energy consumption, thus reducing the probability of the wireless access point from consuming grid power.
\end{enumerate}

\section{Context Aware Traffic Scheduling}
Traffic scheduling reshapes the traffic injected into the network, and hence affects users' data rates, and the spectrum usage and the energy consumption of wireless access networks. Thus, context aware traffic scheduling is desirable in FreeNet to optimize the utilization of spectrum and renewable energy while satisfying the users' QoS requirements.

%
%

\begin{figure*}
\centering
\hspace*{\fill}
       \begin{subfigure}[b]{0.3\textwidth}
      	    \includegraphics[scale=0.2]{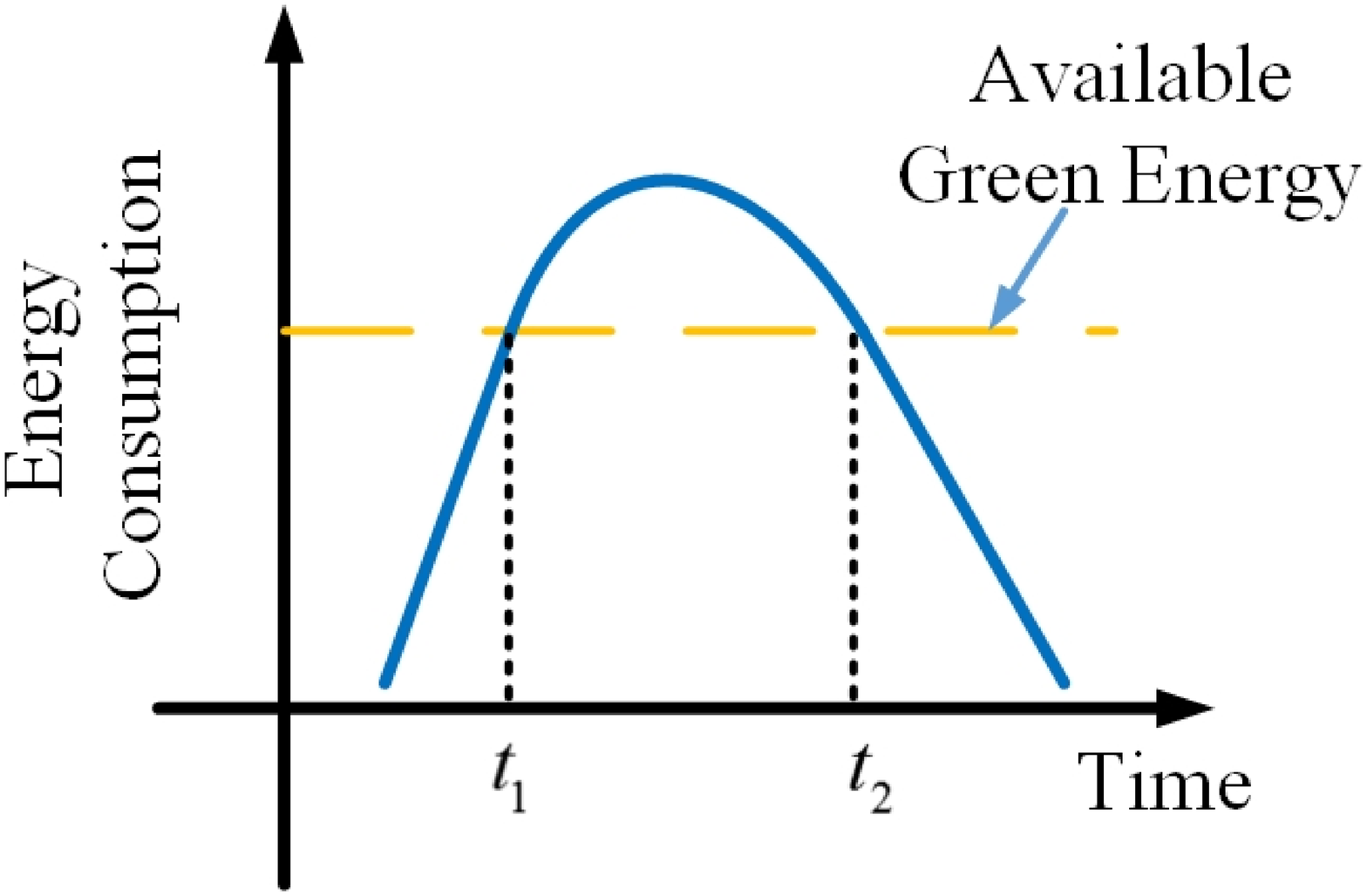}
            \caption{Energy consumption of transmitting incoming traffic.}
            \label{fig:delaya}
       \end{subfigure}\hfill
        \begin{subfigure}[b]{0.3\textwidth}
      	    \includegraphics[scale=0.2]{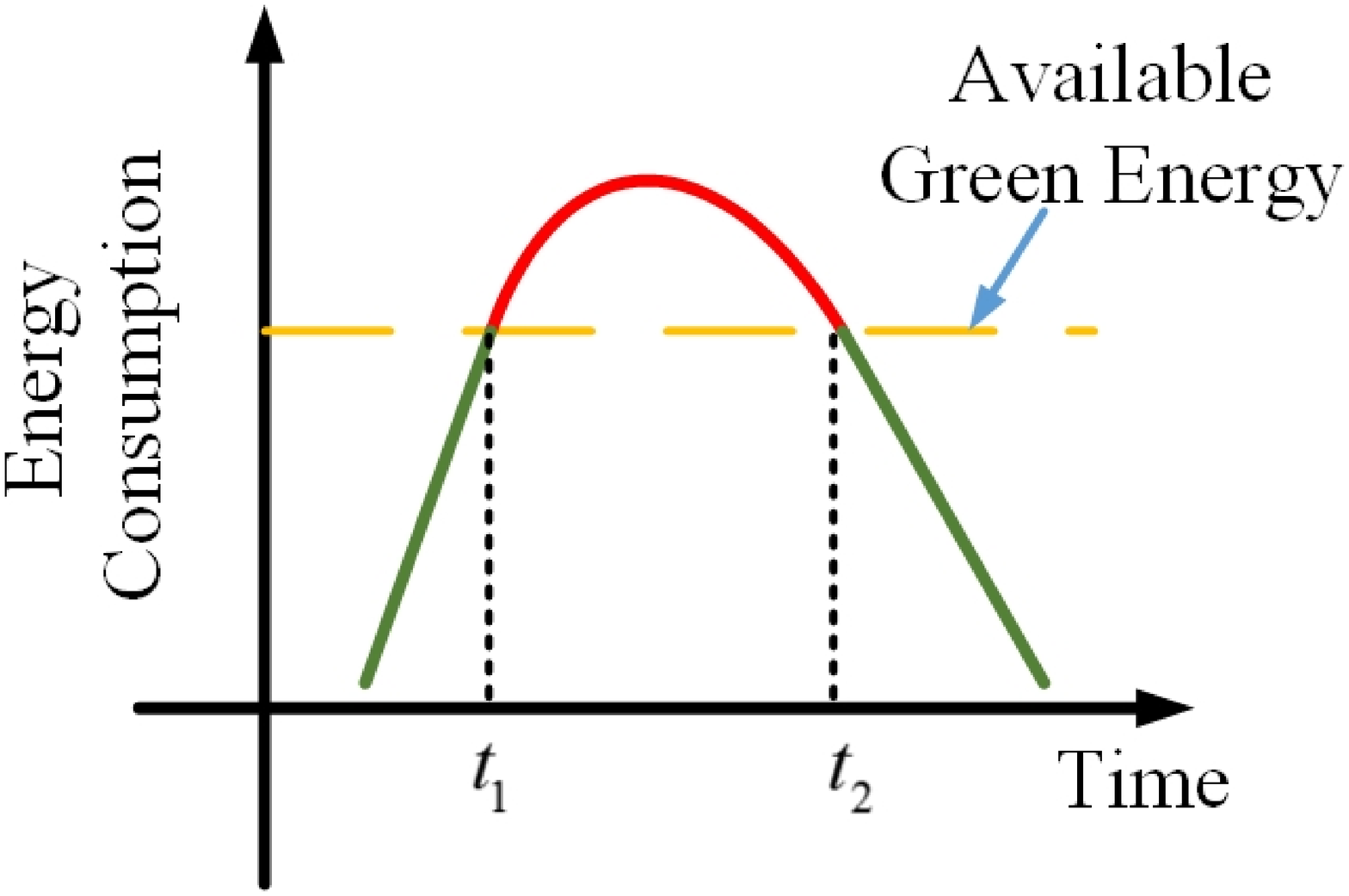}
            \caption{Grid power is drawn (shown in red).}
            \label{fig:delayb}
       \end{subfigure}\hfill
      \begin{subfigure}[b]{0.3\textwidth}
      	    \includegraphics[scale=0.2]{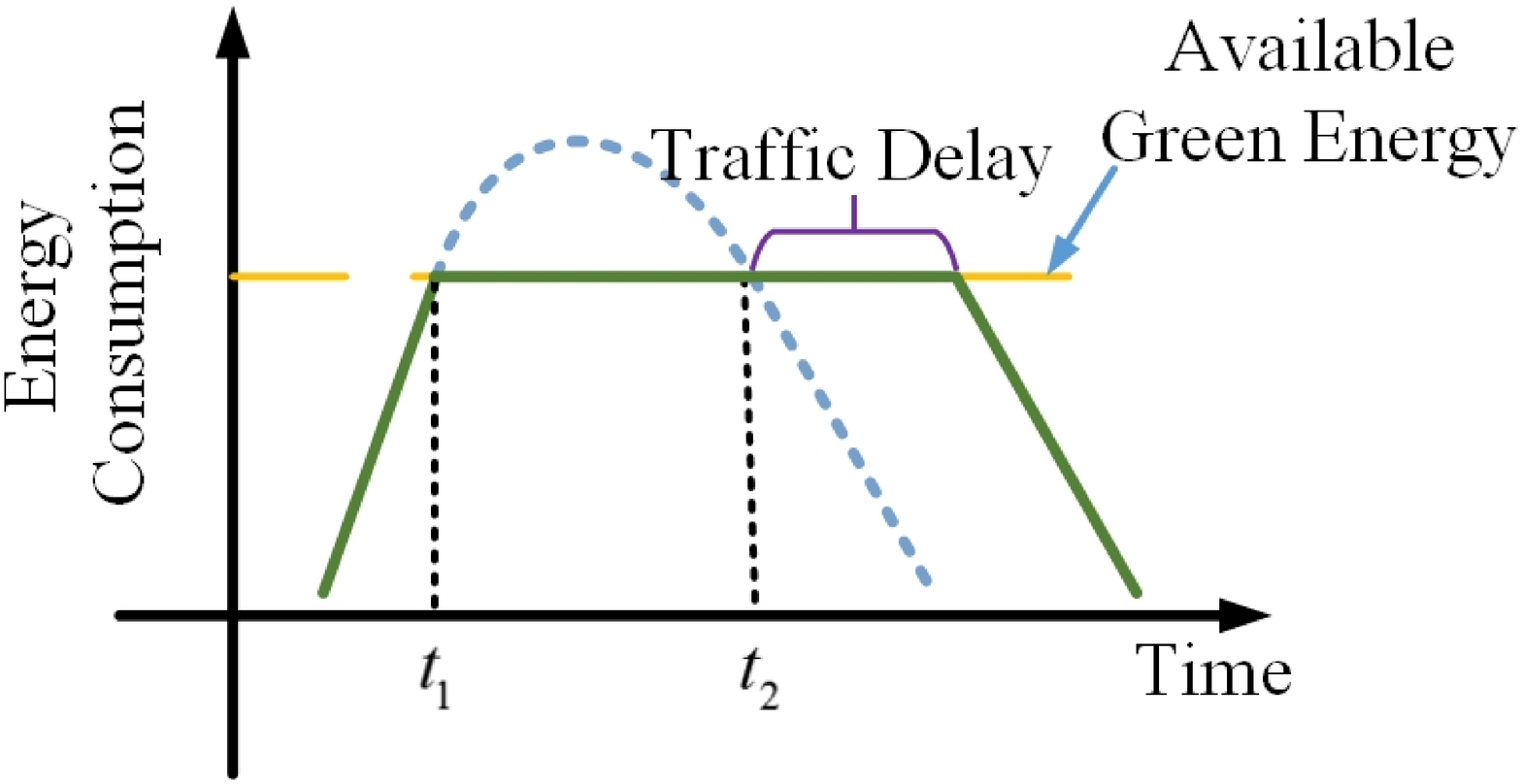}
            \caption{Grid power is avoided by delaying traffic transmission.}
            \label{fig:delayc}
       \end{subfigure}\hfill
    \caption{%
       Grid power consumption v.s. input traffic.
     }%
   \label{fig:delay}
\end{figure*}

\subsection{Grid Power Consumption v.s. Delay}
The power consumption of wireless access points is closely related to their transmission rates. A higher transmission rate usually consumes more energy \cite{Rajan:2004:DPS}. Fig. \ref{fig:delay} shows an example of the trade-off between the grid power consumption and the traffic delay. Assuming the energy consumption is proportional to the traffic transmission rate. Fig. \ref{fig:delaya} shows the energy consumption of transmitting the incoming traffic at its original traffic arrival rate. Fig. \ref{fig:delayb} shows the grid power consumption and the renewable energy consumption when the traffic is transmitted at its arrival rate. When the traffic rate is higher than a threshold, wireless access points require more energy than the available renewable energy, and thus draw energy from the power grid. If the arrival traffic is shaped as shown in Fig. \ref{fig:delayc}, the consumption of the grid power is avoided at the cost of additional packet delay. To guarantee QoS requirements of a user's session, each incoming packet can only delay for a limited duration of time. If the packet delay exceeds the limited duration of time, wireless access points have to utilize grid power to transmit the packet. Therefore, a trade-off between the grid power consumption and packet delay should be determined to minimize the grid power consumption while satisfying users' QoS requirements. In addition, reshaping the outgoing traffic may create the spectrum holes: a wireless access point may not fully utilize its allocated spectrum. In this case, traffic scheduling may be designed to enhance the spectrum efficiency. For example, when a wireless access point seeks additional bandwidth, the other wireless nodes may delay their traffic to create spectrum holes for the wireless access point.

\subsection{Grid Power Consumption v.s. Delay of Diversified Applications}
The above assumes that all the user sessions have the same delay requirement, and traffic of all sessions exhibits the same characteristics. In reality, wireless access networks carry diversified applications which impose different delay requirements and exhibit different traffic characteristics. For example, the legacy telephone service has strict delay requirement and the traffic arrival is constant bit rate (CBR), while the web-surfing service has less stringent delay requirement and the traffic is bursty. For the first kind of traffic, delaying the traffic may not be able to achieve a low average rate, but may exacerbate the delay performance instead, while delaying the traffic of the second class may be able to smooth out the traffic and produce a lower average rate, thus avoiding consuming grid power. It is challenging to design a scheduling algorithm to minimize the grid power consumption when considering these diversified requirements and traffic characteristics.

\subsection{Grid Power Consumption v.s. Loss of Diversified Applications}
In the above analysis, network delay performance is traded for reducing grid power consumption. Loss performance can also be traded for energy saving as well. From the end users' perspective, their quality of experience (QoE) may not be degraded when the packet loss ratio is below some upper bound. Taking the voice application, for example, attributed to the human self error-correction capability, the maximum allowable packet loss ratio can be as high as 3\%. Therefore, properly dropping some traffic may not degrade user service but can help avoid switching the network to grid power supplies. By including the packet loss ratio, the traffic scheduling problem is further complicated, and the challenge of solving the problem is intensified.

\section{Communication Protocol Suite Design}
The above subsections describe the optimal operation strategies on adapting the architecture of access
networks, managing network resources and scheduling traffic for FreeNet. To realize FreeNet, a communication
protocol suite is required to coordinate wireless access points. Designing the communication protocol
suite to enable FreeNet with minimum overhead is challenging. The protocol suite is designed in three different
planes: control plane, energy plane, and user plane.

\subsection{Control plane protocol}
The control plane protocol enables the dynamic network architecture adaptation, spectrum sensing, and spectrum sharing. FreeNet adapts its network architecture according to the dynamics of the available spectrum, renewable energy, and traffic demands. Therefore, under different
conditions, FreeNet may adopt different network operation strategies in terms of network topology, routing algorithms, and transmission techniques. The control plane protocol should be innovated to enable the network architecture adaptation with limited protocol overheads. In addition, FreeNet, by adopting spectrum harvesting techniques, dynamically senses and accesses the available spectrum to enhance the spectrum efficiency of the wireless access network. The control plane protocol should enable wireless
access points to form coalitions to sense the spectrum cooperatively, and allow wireless access points to optimally share the available bandwidth in their coalitions.
\subsection{Power plane protocol}
The power plane protocol facilitates the sharing of the renewable energy information and renewable energy. Since FreeNet is powered by renewable energy, renewable energy is an important network resource which affects the operation of FreeNet. Therefore, the renewable energy information such as energy generation rates, renewable energy consumption, and renewable energy storage should be shared among the wireless access points in order to optimize the network architecture and network resource allocations. Thus, the power plane protocol is required to share the renewable energy information among wireless access points, and control the renewable energy sharing and distribution among wireless access points to enhance the utilization of renewable energy.
\subsection{User plane protocol}
The user plane protocol is responsible for data traffic scheduling, packet routing, and user handover among various types of wireless access points. Traffic scheduling is essential to yield the optimal trade off among the grid power consumption, packet delay and packet loss. In order to optimize traffic scheduling, wireless access points should be aware of the renewable energy and traffic characteristics. The user plane protocol is proposed to enable wireless access points to obtain such information for traffic scheduling. In addition, FreeNet consists of both infrastructure-based wireless access networks and mesh-based wireless access networks. The user plane protocol should enable packet routing via heterogeneous networking, and allow handover between two networks with the minimal packet delay and packet loss.

\section{Conclusion}
This article introduces FreeNet that liberates the spectrum and energy constraints of the wireless networks by exploiting spectrum and energy harvesting technologies. We have discussed the deployment scenarios of FreeNet including alleviating network congestion in urban areas, provisioning broadband services in rural areas, and upgrading emergency communication capability in critical situations. Since the radio spectrum and renewable energy are highly dynamic, it is challenging to design FreeNet. In this article, we have briefly analyzed the related research issues on network architecture optimization, network resource management, and the context aware traffic scheduling. These discussions shed light on designing and optimizing future spectrum and energy efficient wireless networks.
\bibliographystyle{IEEEtran}
\bibliography{mybib}

\begin{IEEEbiography}
[{\includegraphics[width=1in,height=1.25in,clip,keepaspectratio]{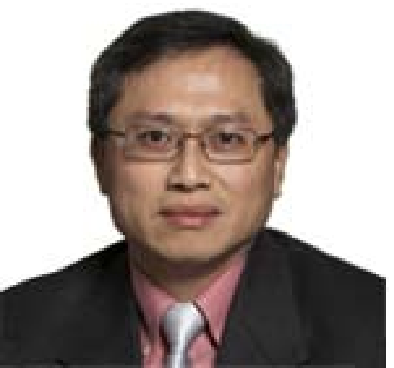}}]{Prof. Ansari}[S’78-M’83-SM’94-F’09] received the B.S.E.E. degree (summa cum laude) from the
New Jersey Institute of Technology (NJIT), Newark, NJ, USA, the M.S.E.E. degree from the University of Michigan, Ann Arbor, MI, USA, and the Ph.D. degree from Purdue University, West Lafayette, IN, USA.

He is a Distinguished Professor of Electrical and Computer Engineering at NJIT, which he joined in 1988. He has also assumed various administrative positions at NJIT. He has been a Visiting (Chair) Professor at several universities. He coauthored Media Access Control and Resource Allocation (Springer, 2013) with J. Zhang and Computational Intelligence for Optimization (Springer, 1997) with E.S.H. Hou, and edited Neural Networks in Telecommunications (Springer, 1994) with B. Yuhas. He has also contributed over 450 technical papers, over one-third of which were published in widely cited refereed journals/magazines. He has guest edited a number of Special Issues, covering various emerging topics in communications and networking. His current research focuses on various aspects of broadband networks and multimedia communications.

He has served on the Editorial Board and Advisory Board of nine journals, including as a Senior Technical Editor of IEEE Communications Magazine (2006–2009). He was elected to serve on the IEEE Communications Society (ComSoc) Board of Governors as a Member-at-Large (2013-2015). He has
chaired ComSoc Technical Committees, and has actively organized numerous IEEE international conferences/symposia/workshops, assuming various leadership roles. Some of his recognitions include several Excellence in Teaching Awards, two Best Paper Awards, the NCE Excellence in Research Award
(2014), ComSoc AHSN TC Outstanding Service Recognition Award (2013), NJ Inventors Hall of Fame Inventor of the Year Award (2012), Thomas Alva Edison Patent Award (2010), and designation as an IEEE Communications Society Distinguished Lecturer (2006-2009, two terms). He has also been
granted over 20 U.S. patents.
\end{IEEEbiography}
\begin{IEEEbiography}
[{\includegraphics[width=1in,height=1.25in,clip,keepaspectratio]{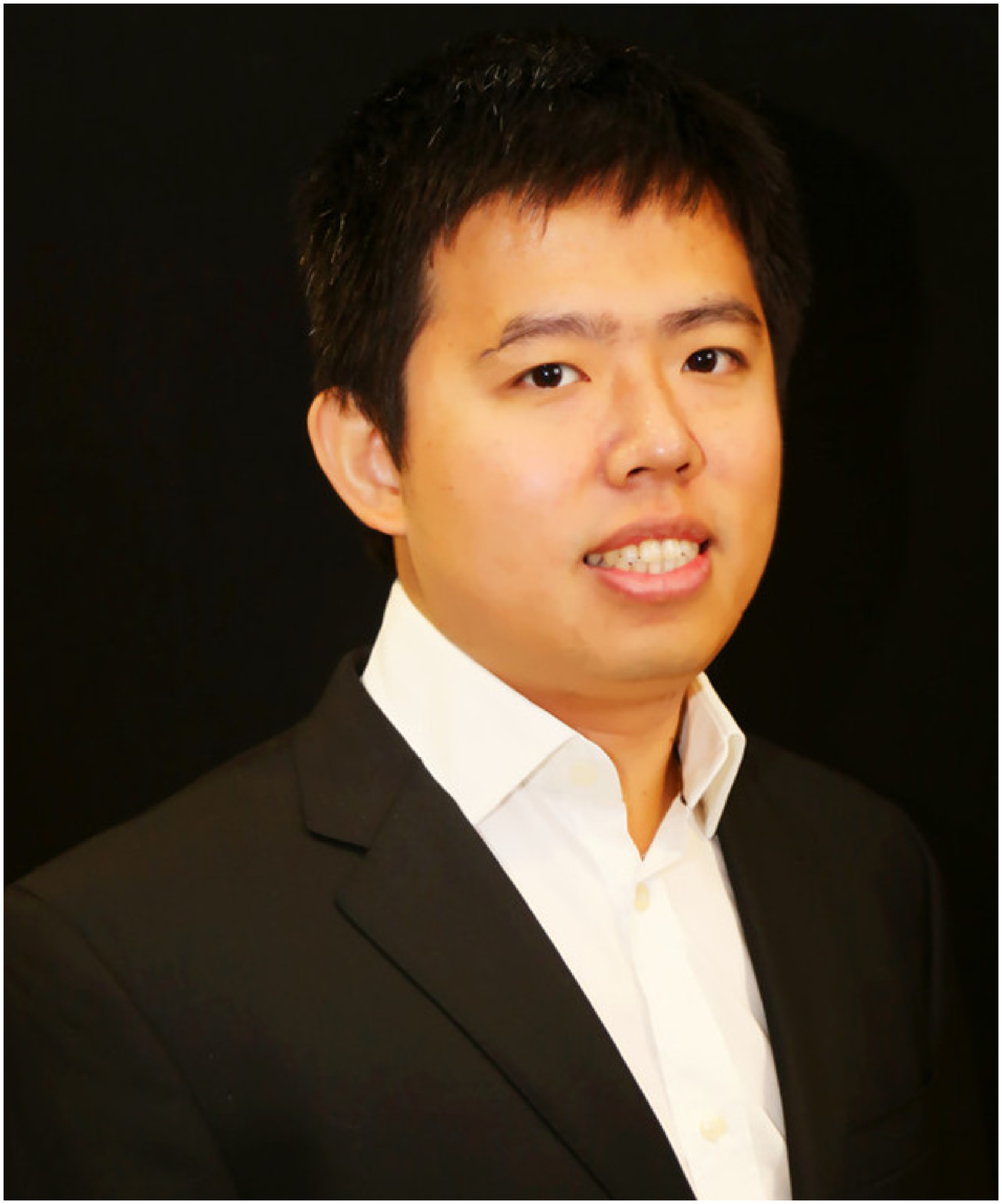}}]{Tao Han} [S'08] is a Ph.D. candidate in the Department of Electrical and Computer Engineering at the New Jersey Institute of Technology (NJIT),
Newark, NJ, USA. His research interests are big-data-driven communication network design, mobile and wireless networking, the Internet of
Things, and green communications. He has authored 11 papers published/accepted in premium IEEE publications such as ACM/IEEE Transactions on Networking, IEEE Transactions on Wireless Communications, and IEEE Wireless Communications. He has also produced 13 high-quality IEEE conference papers and five U.S. non-provisional patent applications. He has been recognized with a New Jersey Inventors Hall of Fame Graduate Students Award.
\end{IEEEbiography}

\end{document}